\documentclass[numberedappendix,apj]{emulateapj}

\usepackage{apjfonts}

\begin{document}

\shorttitle{Entrainment in jet+sidewind systems}

\shortauthors{L\'opez-C\'amara \& Raga}

\title{Side-entrainment in a jet embedded in a sidewind}

\author{D. L\'opez-C\'amara\altaffilmark{1}, A. C. Raga\altaffilmark{1}}
\email{diego.lopez@nucleares.unam.mx}

\altaffiltext{1}{Instituto de Ciencias Nucleares, Universidad
Nacional Aut\'onoma de M\'exico, Ap. 70-543,
04510 D.F., M\'exico}

\begin{abstract}
Numerical simulations of HH jets never show side-entrainment of
environmental material into the jet beam. This is because the bow
shock associated with the jet head pushes the surrounding environment
into a dense shell, which is never in direct contact with the
sides of the jet beam. We present 3D simulations in which a
side-streaming motion (representing the motion of the outflow source
through the surrounding medium) pushes the post-bow shock shell into
direct contact with the jet beam. This is a possible mechanism for
modelling well collimated ``molecular jets'' as an atomic/ionic
flow which entrains molecules initially present only in the
surrounding environment.
\end{abstract}


\section{Introduction}
\label{sec:int}

Models of entrainment of molecular, environmental material, in the
wings of bow shocks (associated with working surfaces in jets
from young stars), are successful in explaining the limb-brightened,
cavity-like molecular outflows (at least in a qualitative way).
Analytic \citep{mas93,rc93} and numerical \citep{rag95,lim01}
models of this so-called ``prompt entrainment'' scenario produce
limb-brightened molecular structures that resemble the
cavity-lile morphologies observed in objects such as the
L1157 outflow \citep[e. g.,][]{bel04}.

However, some outflows from young stars also show high-velocity,
collimated, jet-like molecular structures. An example
of this kind of structure is observed in HH212 \citep[e. g.,][]{cod07}.
These jet-like molecular outflows have been successfully modeled by
assuming that the jet itself is initially molecular
\citep[e. g.,][]{vol99,lim01,mor06}.

Could these ``molecular jets'' be the result of environmental
molecular gas being entrained into an atomic/ionic jet? The
possibility of having ``side entrainment'' of molecular material
has been studied analytically \citep{can91} and numerically
\citep{tay95,lim99}. These models show that if one has a
fast, atomic jet beam in direct contact (through the sides
of the beam) with a molecular environment,
a substantial amount of molecular material is indeed entrained into
the fast flow.

However, if one computes full simulations of a jet flow, the leading
bow shock pushes aside the molecular environment, so that the
sides of the jet beam are never in direct contact with the molecular
gas. Therefore, the situation necessary for producing
side-entrainment of molecular gas \citep[see][]{tay95,lim99} is
not obtained.

In the present paper, we study the possibility of overcoming
this problem by having a low velocity side-motion of the environment
relative to the jet source. This side-wind could represent the
motion of the jet source within the surrounding environment.
The qualitative effect of the sidewind is described in \S 2.
We have then computed a set of 3D simulations of a radiative jet in a
sidewind (\S 3), producing a variety of flow morphologies (\S 4).
From the resulting flows we compute the amount of environmental
material which is pushed by the jet flow (\S 5) and analyze
how much material is actually entrained into the jet beam itself
(\S 6). We finally illustrate the dependence of our results
on the resolution of the numerical simulations (\S 7). The
results are summarized in \S 8.

\section{Jet in a sidewind}
\label{sec:jetwind}

\begin{figure}[!h]
\epsscale{0.8}
\plotone{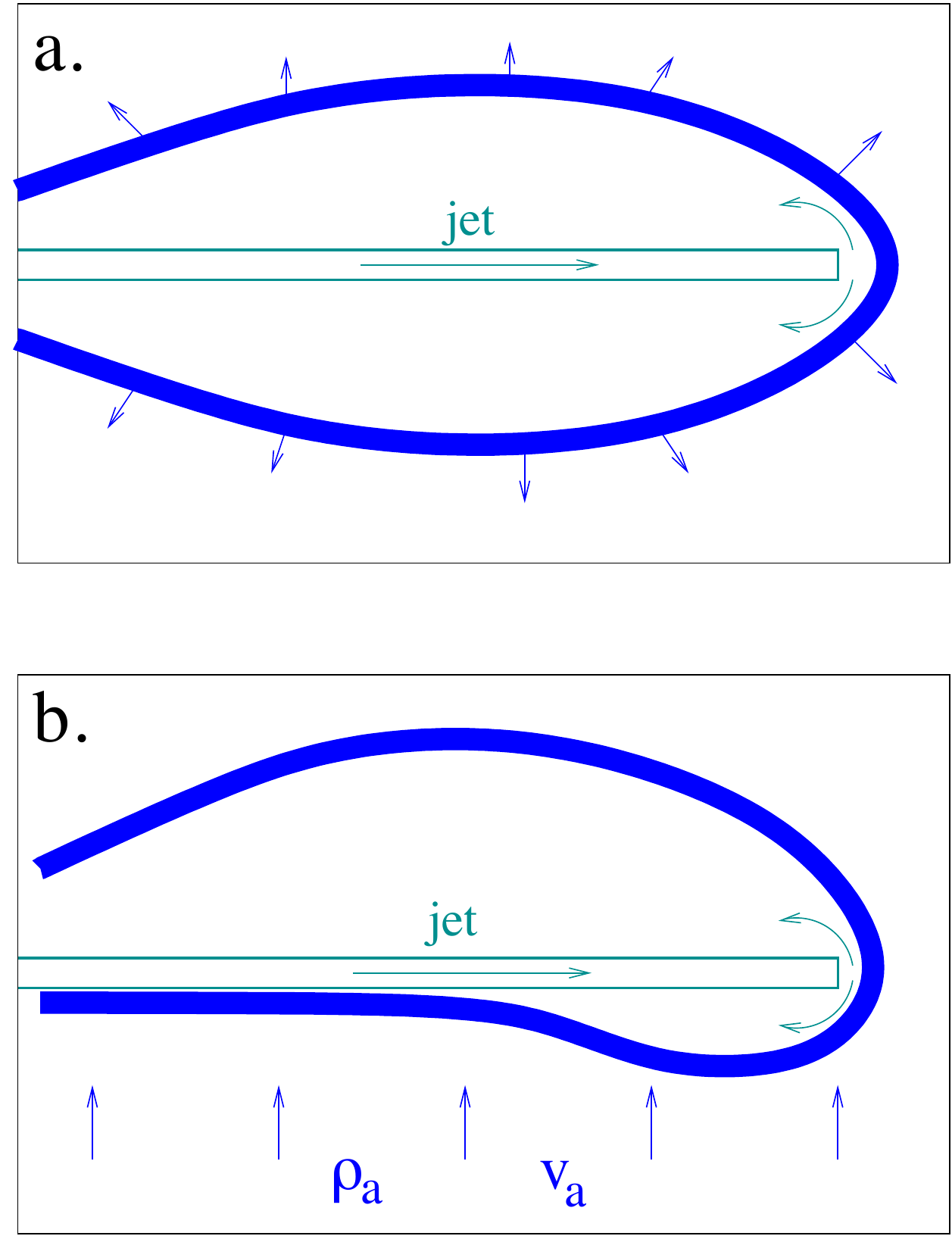}
\caption{Schematic diagram showing a jet travelling in
a stationary environment (top) and in a sidewind travelling
parallel to the ordinate (bottom). The sidewind pushes the
dense, post-bow shock shell into direct contact with the
jet beam.}
\label{fig1}
\end{figure}

The main problem when trying to incorporate molecular,
environmental material, into a collimated jet is that the
leading head of the jet, and possibly also any trailing
``internal working surfaces'', push away the environmental
gas into a dense shell, which follows the shape of the
bow shock wings. Because of this, the molecular environmental
material never reaches contact with the jet beam, and
lateral entrainment of this material into the jet does
not occur. This situation is shown in panel a of Figure~\ref{fig1}.

It has been suggested that if the dense shell material were
warm enough, it might re-expand into the cavity left by
the passage of the jet head, and reach contact with the
jet beam \citep[see, e.~g.,][]{rc93}. However, this is found
not to be an important effect in jet simulations
\citep[see, e.g.,][]{rag95,lim01}.
Possibly, a stratification of the
surrounding environment and/or a precession and
variability of the jet ejection could lead to
the occurrence of side-entrainment into the jet beam \citep{cab97}.
However, ntil now the correct combination of parameters
for this to occur has not been found. Other possibilities have been
suggested. For example, \citet{lim01} studied the survival of
molecules (originally present in the environment) in the head of
an accelerating jet flow. It is not clear how this
high-velocity molecular gas in the jet head could end up being
entrained into the jet beam. Another possibility was
suggested by \citet{rag03},
who proposed that the existence of small, dense, moving clumps
within the environment might be a way of introducing molecular
material into the jet beam. However, it is not clear that
this mechanism leads to molecular emission structures that
resemble the observations. 

The possibility that we study in this paper is that the
presence of a side-streaming environment pushes the bow shock
wing (and the post-bow shock, dense shell) against the jet
beam, as shown in panel {\it b.} of Figure~\ref{fig1}. The side-streaming
could be the result of the motion of the jet source through
the molecular cloud, and would have velocities of at most a
few km~s$^{-1}$. Both analytic \citep{can95,rag09} and numerical
\citep{lim98,mas01,cia08} models of jets in sidewinds have been
computed previously. These models have mostly been applied to
jets within expanding
H~II regions \citep{cia08} or to jets embedded
in an isotropic stellar wind \citep{rag09}. The relatively
high sidewind velocities relevant for these cases ($\sim 10$-30~km~s$^{-1}$
for an expanding H~II region and up to $\sim 1000$~km~s$^{-1}$
for a stellar wind), can produce jets with strongly curved jet beams.

For lower sidewind velocities (not studied in the papers cited
above), only a weak curvature will be produced
as a result of the jet/sidewind interaction. However, the bow shock
wing will still be pushed against the body of the jet, as shown in
the schematic diagram of Figure~\ref{fig1}. We focus on this regime, in which
the dense shell of swept up environmental material is pushed into contact
with the body of the jet beam (therefore allowing the entrainment of
environmental gas into the beam), but a relatively straight jet path
is still obtained. The numerical simulations which we have carried out
are described in the next section.

\section{The numerical simulations}
\label{sec:num}

We have computed a set of 3D gasdynamic simulations of jet/sidewind
interactions. All of them have been computed in a $(2,0.5,0.5)\times
10^{17}$~cm cartesian grid. The jet is injected at $x=0$ (in the
centre of the boundary plane of the computational grid), with
a velocity parallel to the $x$-axis. A sidewind
is injected in the $y=0$ plane, with a velocity directed
along the $y$-axis. A reflection boundary condition is applied
on the $x=0$ boundary outside the jet beam, and transmission conditions
are applied on all of the other boundaries except the $y=0$ plane (in which
the sidewind is injected).

An initially neutral, top-hat jet of velocity $v_j$, density $n_j$, radius
$r_j=10^{15}$~cm and temperature $T_j=10^3$~K moves into an initially
uniform, neutral environment with a density $n_a=200$~cm$^{-3}$,
temperature $T_a=10$~K and sidestreaming velocity $v_a$.
A set of models with different values of $v_j$, $n_j$ and $v_a$ has
been computed, with the parameters given in Table~1.

\begin{table}\centering
\caption{Model characteristics.}
\label{tab1}
\begin{tabular}{ccccc}
\hline
\hline
Model & $v_j$ & $v_a$ & $n_j$ \\
      & [km s$^{-1}$] & [km s$^{-1}$] & [cm$^{-3}$] & resolution \\
\hline
{\bf a1} & 150 & \phantom{0}2 & 1000 & lr,mr,hr\\
{\bf b1} & 150 & \phantom{0}5 & 1000 & mr \\
{\bf c1} & 150 & 10 & 1000 & lr,mr,hr \\
{\bf a2} & 300 & \phantom{0}2 & 1000 & mr \\
{\bf b2} & 300 & \phantom{0}5 & 1000 & mr \\
{\bf c2} & 300 & 10 & 1000 & mr \\
{\bf a3} & 150 & \phantom{0}2 & 5000 & mr \\
{\bf b3} & 150 & \phantom{0}5 & 5000 & mr \\
{\bf c3} & 150 & 10 & 5000 & mr \\
\hline 
\hline 
\end{tabular}
\end{table} 

The simulations were carried out with the ``Yguaz\'u-a'' code \citep{rag00}, solving the 3D gasdynamic equations together with a continuity/rate equation for neutral H. The parametrized cooling function described by \cite{rag09b} is included in the energy equation. Also, we integrate an equation for a normalized passive scalar $g$ with which we distinguish between the ambient and jet medium. If the scalar was positive it indicated that the material was initially medium material, while if it was negative it was jet material. For example, if we had only ambient medium material $g=1$, or if we had pure jet material $g=-1$.
With the use of this scalar we were able to calculate the amount of mixing between the ambient medium and the jet material. For this, we defined the mixing mass fraction as $f_m = (g+1)/2$. It is clear that for pure ambient medium $f_m=1$; while for pure jet material $f_m=0$, intermediate values of $f_m$ indicates that there was material from both the ambient medium and the jet mixed together. The case in which we have 99\% ambient medium mixed with 1\% of jet material ($f_m=0.99$), is defined as ``99\% mixing fraction''; while the 1\% ambient medium case ($f_m=0.01$) as ``1\% mixing fraction''.

A 6-level binary adaptive grid has been used with three different maximum resolutions $\Delta x = 3.91$, 1.95 and $0.98\times 10^{13}$~cm (along the three axes).
We have called these the ``low'', ``medium'' and ``high'' resolutions
(labeled in Table~1 with letters lr, mr and hr, respectively).

\section{The resulting flow stratifications}
\label{sec:flow}

\begin{figure}[!h]
\epsscale{1.25}
\plotone{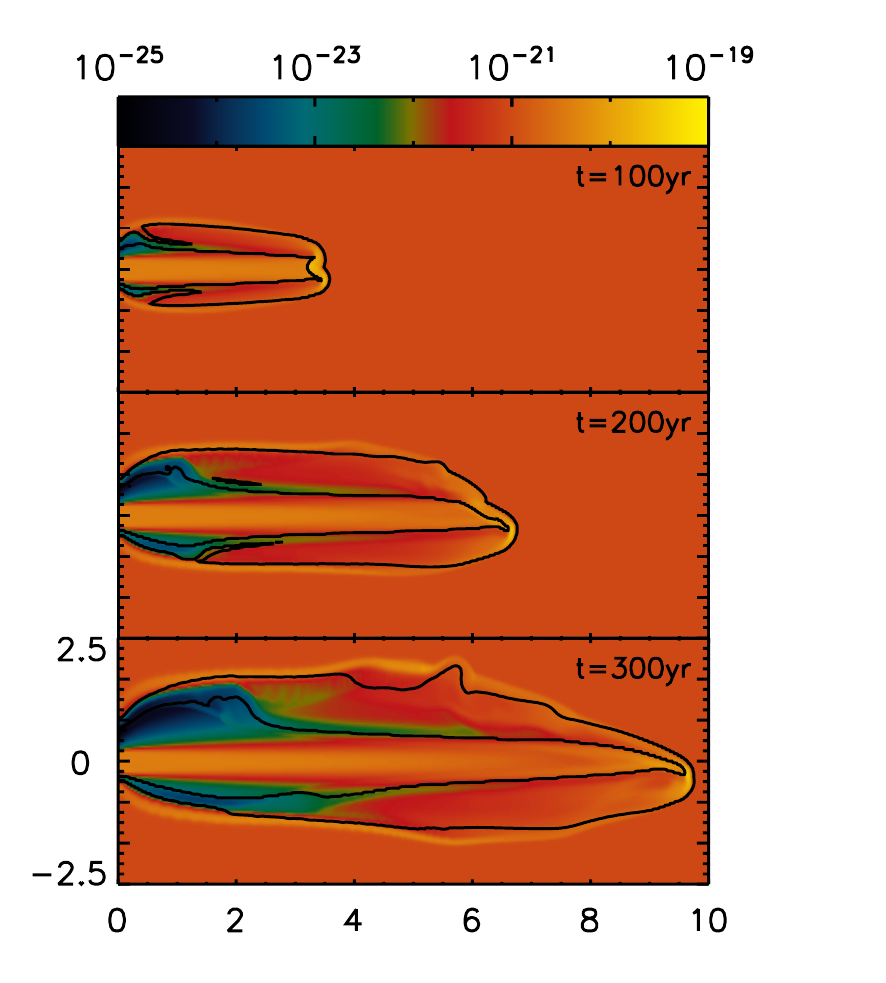}
\caption{Density stratifications (color scale, given in g cm$^{-3}$ by the top bar) and mixing fractions (contours) obtained from model a1 (see Table 1). The side-streaming environment flows along the vertical axis of the plots. The displayed stratifications correspond to cuts on a plane that includes the outflow axis and the sidewind, at integration times $t=100$ (top), 200 and 300~yr (bottom). The two contours correspond to the 99~\%\ mixing fraction ($f_m=0.99$, outer contour); and the 1~\%\ mixing fraction case ($f_m=0.01$, inner contour). The axes are labeled in units of $10^{16}$~cm.}
\label{fig2}
\end{figure}

As an example of the flows resulting from our simulations, in Figure~\ref{fig2} we show $xy$-cuts showing the time-evolution of the mid-plane density stratification obtained from model a1 (see Table~1). This figure also shows 2 contours, corresponding to the 99\% mixing fraction ($f_m=0.99$, outer contour), and 1\% mixing fraction ($f_m=0.01$, inner contour).

In model a1, the $xy$-cuts show a side-to-side asymmetry that
is a direct result of the fact that the environment is flowing
along the $+y$-direction. This asymmetry is seen as a distortion
of the leading bow shock, and as a penetration of environmental
material to regions close to the jet beam in the up-sidewind
($-y$) direction. 

\begin{figure*}[!t]
\epsscale{1.1}
\plotone{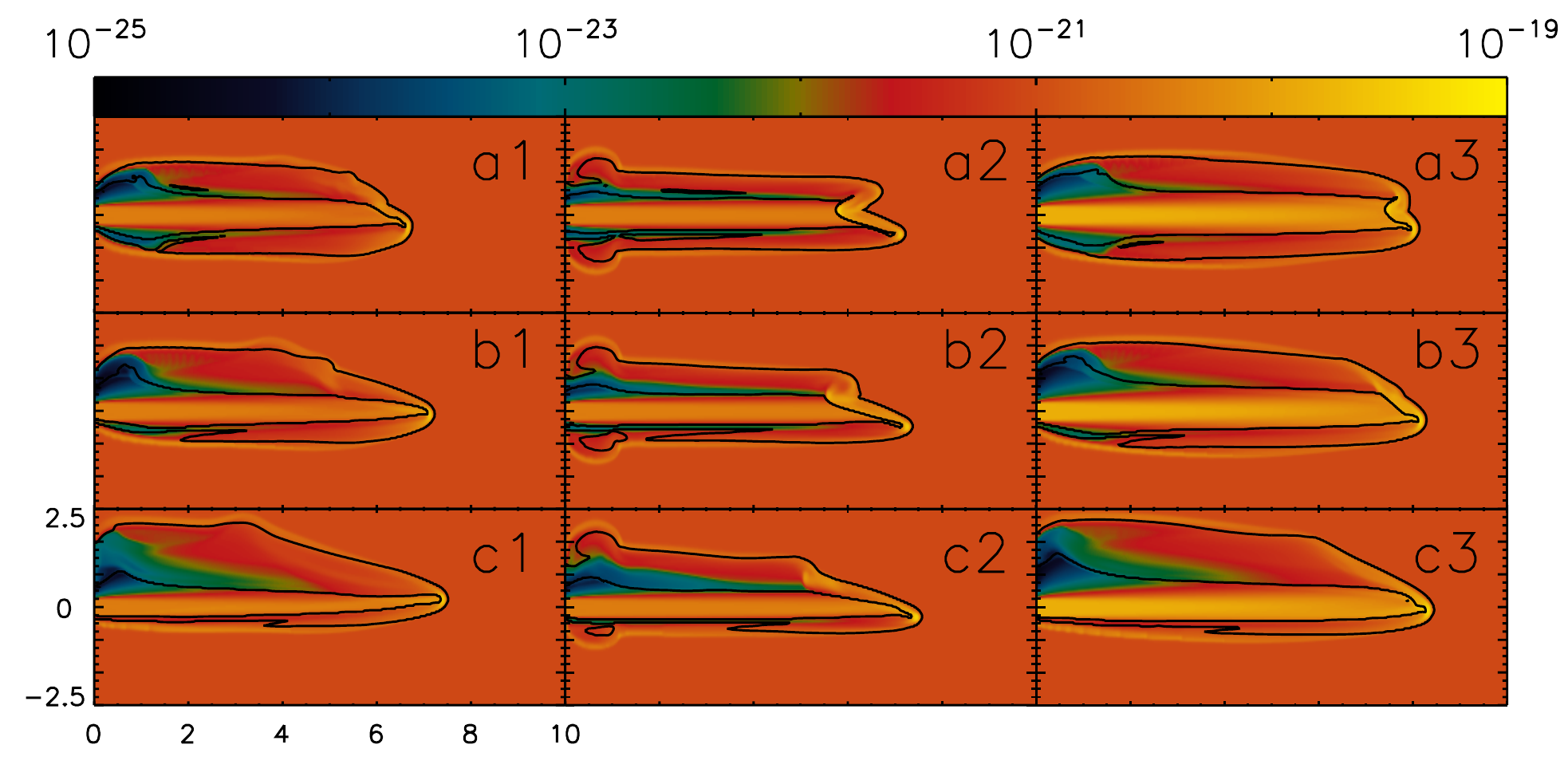}
\caption{Density stratifications (color scale, given in g cm$^{-3}$ by the top bar) and mixing fractions (contours) obtained for the models listed in Table 1. The displayed stratifications correspond to cuts on a plane that includes the outflow axis and the sidewind, at integration times $t=200$~yr (for models a1, b1, c1, a3, b3 and c3) or $t=100$~yr (for models a2, b2, c2). The two contours correspond to the 99~\%\ mixing fraction ($f_m=0.99$, outer contour); and the 1~\%\ mixing fraction case ($f_m=0.01$, inner contour). The axes are labeled in units of $10^{16}$~cm.}
\label{fig3}
\end{figure*}

In Figure~\ref{fig3}, we show single time frames of the $xy$-midplane
density stratifications obtained from the 9 models of Table~1. In each
of the three columns of Figure~\ref{fig3} (from top to bottom), we see the
effect of increasing velocity of the sidewind (the top,
centre and bottom frames correspond to $v_a=2$, 5 and 10~km~s$^{-1}$,
respectively, see Table~1). It is clear that for higher values
of $v_a$, the environmental material penetrates more
strongly towards the jet beam in the up-sidewind region.

A comparison of the first and second columns of Figure~\ref{fig3}
shows that if one increases the jet velocity from
$v_j=150$~km~s$^{-1}$ (models
a1, b1 and c1) to 300~km~s$^{-1}$ (models a2, b2 and c2,
see Table 1) the resulting density stratifications and
mixing fractions remain qualitatively unchanged,
showing slightly more pronounced asymmetries for
the higher jet velocity. In particular, it is clear
that in regions close to the source
the post-bow shock dense shell touches the
jet beam in all of the $v_a=10$~km~s$^{-1}$ models (bottom row
of Figure~\ref{fig3}).

Finally, we also see that
if one increases the jet density (from $n_j=1000$~cm$^{-3}$
for the models in the first column to 5000~cm$^{-3}$
for the models in the third column, see Table 1),
less penetration of the environmental material
into the jet beam is obtained. This effect can be seen as
broader 99\%\ pure jet material regions (limited by the
inner contour) in the a3, b3, c3 models (compared to a1, b1 and c1).

\section{Entrained material}
\label{sec:ent}

As we have described in \S\ref{sec:num}, from our simulations
we obtain the environmental to total mass mixing fraction $f_m$
as a function of position and time. Using this mixing fraction,
we compute the jet mass loss rate associated with motions
along the $x$-axis~:
\begin{equation}
{\dot M}_j(x)=\int\int \left[1-f_m\right]\rho u\, dy\, dz\,,
\label{mj}
\end{equation}
and the mass rate of the entrained material
\begin{equation}
{\dot M}_{AM}(x)=\int\int f_m \rho u\, dy\, dz\,,
\label{mam}
\end{equation}
where $f_m$, $\rho$ and $u$ are the 3D mixing fraction, density
and $x$-velocity (respectively), obtained for a
given integration time $t$.

\begin{figure}[!h]
\epsscale{1.0}
\plotone{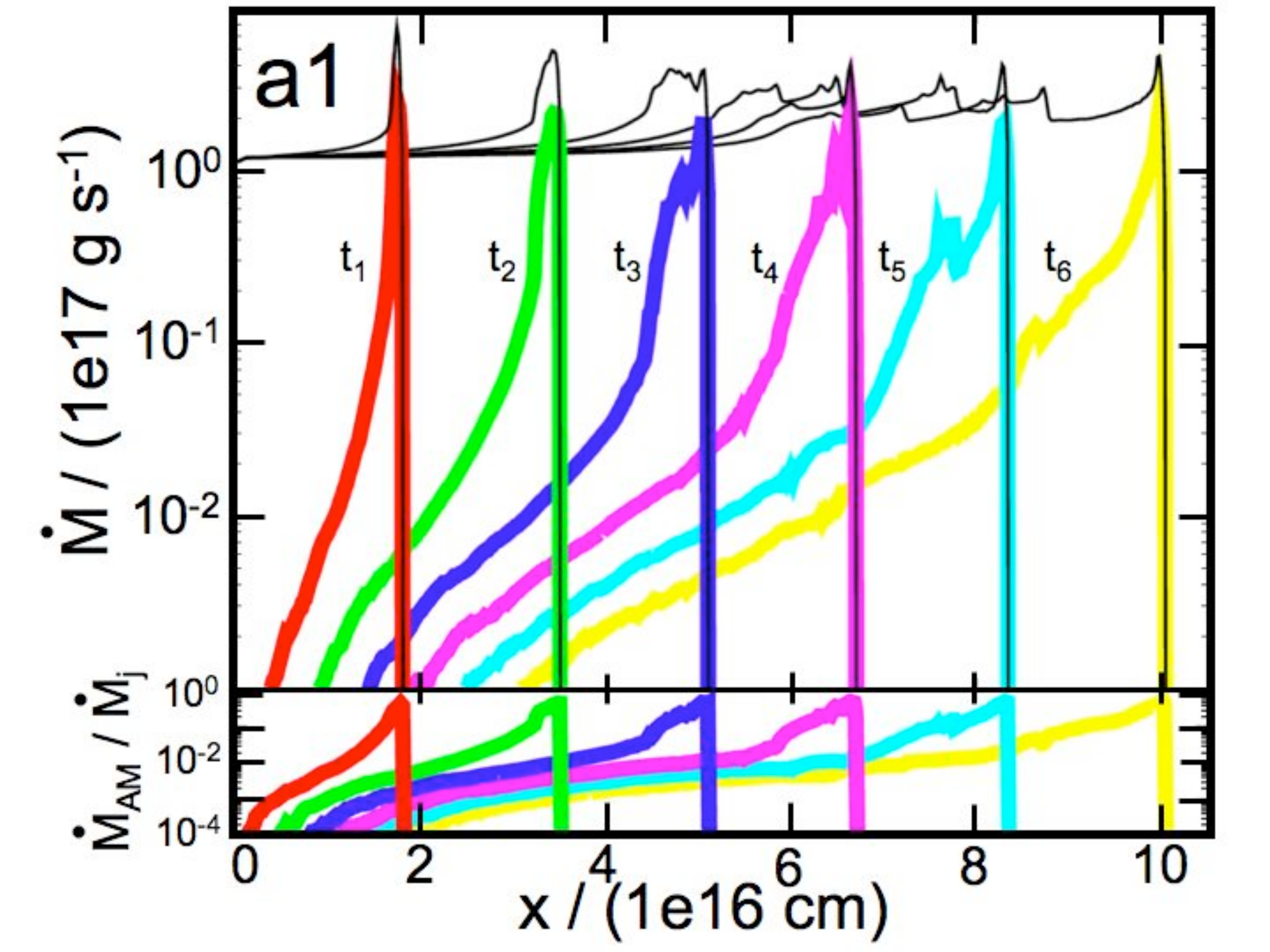}
\caption{Top panel:
Mass rates associated with the jet (${\dot M}_j$,
black lines) and environmental material (${\dot M}_{AM}$,
coloured lines) as a function
of position $x$ along the outflow axis, obtained from
the flow stratifications of model a1 (see Table 1) at
integration times $t_1=50$ (red lines), $t_2=100$ (green),
$t_3=150$ (blue), $t_4=200$ (pink), $t_5=250$ (cyan) and
$t_6=300$~yr (yellow). Bottom panel: the
${\dot M}_{AM}/{\dot M}_j$ ratio as a function of $x$ for
the same integration times.}
\label{fig4}
\end{figure}

\begin{figure}[!h]
\epsscale{1.0}
\plotone{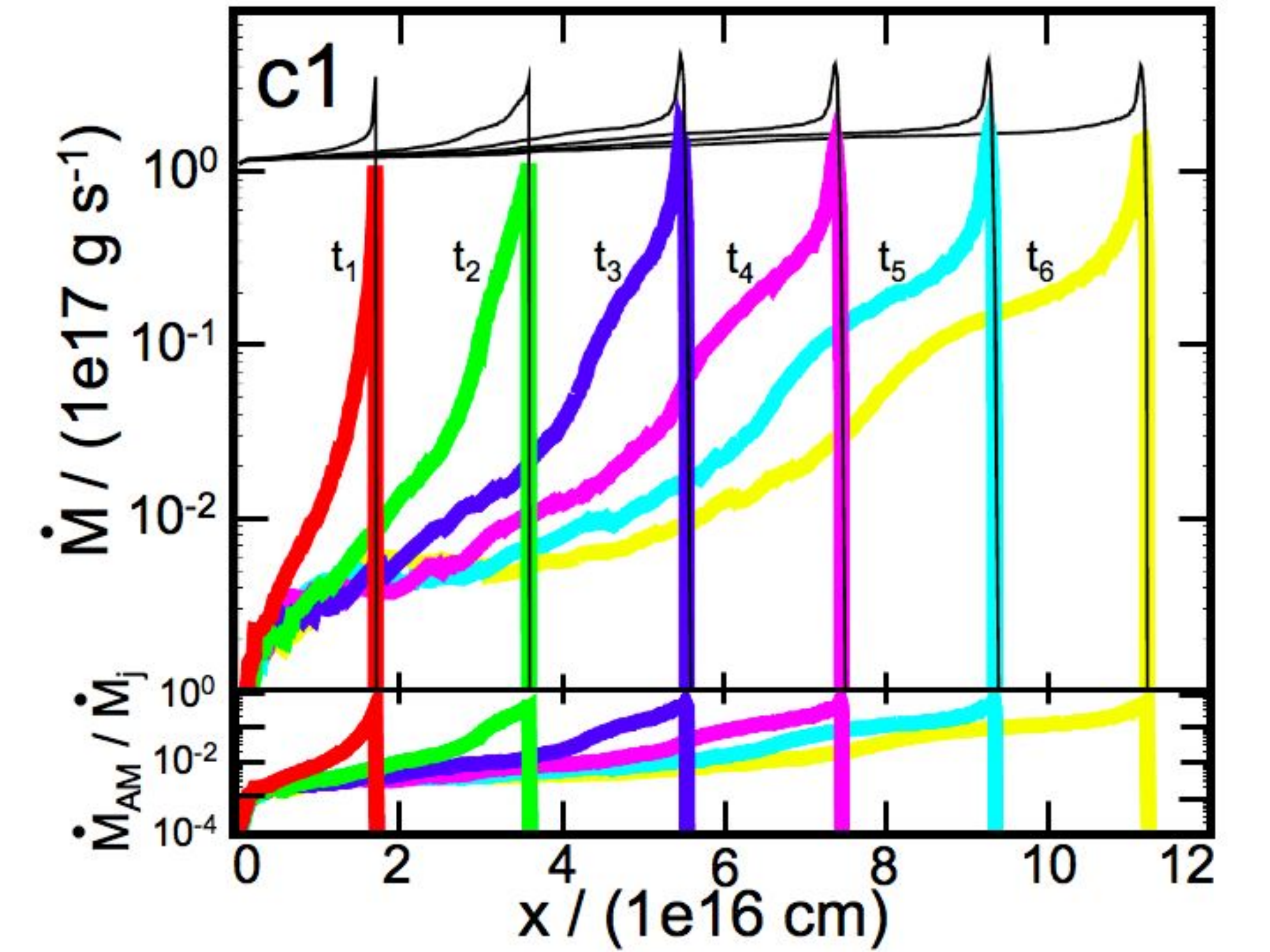}
\caption{The same as Figure 4, but for model c1 (see
Table 1).}
\label{fig5}
\end{figure}

The mass loss rates obtained in this way for models a1 and c1 are shown in Figure~\ref{fig4} and \ref{fig5} (for times $t=50$ to 300~yr). It must be noted that calculating the position of the jet head was not trivial, it required fine-tuning, was not obtained exactly the same for each model, and the environmental mass rate from the entrained material (${\dot M}_{AM}$) was extremely sensitive to it. Thus, in order to be consistent in our analysis, we exclude the head of the jet from our discussion. For model a1 (which has a $v_a=2$~km~s$^{-1}$ sidewind, see Table 1), we see that, at all times ${\dot M}_{AM}$ monotonically grows along the jet axis, and that a maximum is reached at the position of the jets head (top panel of Figure~\ref{fig4}). For this model, the ${\dot M}_{AM}/{\dot M}_j$ ratio also grows with distance $x$ from the source, having values of $\sim 5\to 8\times 10^{-3}$ at the middle of the length of the jet (at a given integration time), and reaching values of $\sim 1$ at the head of the jet (bottom panel of Figure~\ref{fig4}).

For model c1 (which has the same parameters as model a1 except for a $v_a=10$~km~s$^{-1}$ sidewind, see Table 1), a qualitatively similar behavior is obtained for the ambient mass rate, but with higher values of ${\dot M}_{AM}$ at all times and positions along the jet (top panel of Figure~\ref{fig5}). The ${\dot M}_{AM}/{\dot M}_j$ ratio has values of $\sim 1.1\to 1.3\times 10^{-2}$ at the middle of the length of the jet (for all times), a factor of $\sim 2$ larger than the values obtained for model a1 (also except for the head of the jet, which we excluded from the analysis due to lack of consistency).

We now calculate the mean velocities associated with
the jet and environmental mass rates~:
\begin{equation}
\overline{v}_j(x)=\frac{1}{{\dot M}_j(x)}
\int\int \left[1-f_m\right]\rho u^2\, dy\, dz\,,
\label{vj}
\end{equation}
\begin{equation}
\overline{v}_{AM}(x)=\frac{1}{{\dot M}_{AM}(x)}
\int\int f_m\,\rho u^2\, dy\, dz\,,
\label{vam}
\end{equation}
where ${\dot M}_j(x)$ and ${\dot M}_{AM}$ are
given by equations (\ref{mj}) and (\ref{mam}).

\begin{figure}[!h]
\epsscale{1.1}
\plotone{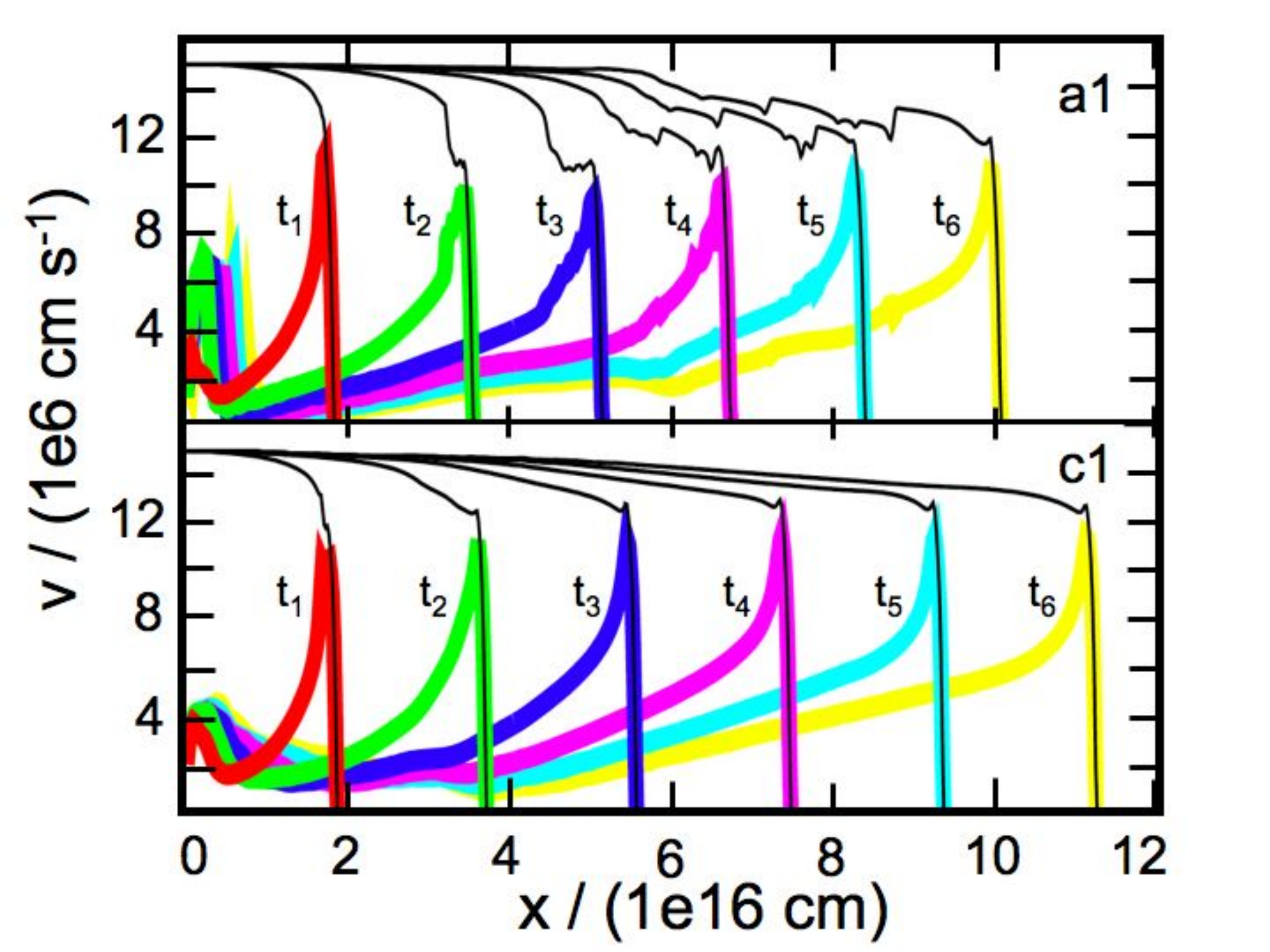}
\caption{Top panel:
Average velocity associated with the jet ($\overline{v}_{j}$, black lines) and the environmental material ($\overline{v}_{AM}$, coloured lines) as a function of position $x$ along the outflow axis, obtained from the flow stratifications of model a1 (see Table 1) at integration times $t_1=50$ (red lines), $t_2=100$ (green), $t_3=150$ (blue), $t_4=200$ (pink), $t_5=250$ (cyan) and $t_6=300$~yr (yellow).
Bottom panel: 
The same as in the top panel, but for model c1 (see Table 1).}
\label{fig6}
\end{figure}

The mean velocities obtained in this way for models
a1 and c1 are shown in Figure~\ref{fig6}. We see that the
mean forward velocity $\overline{v}_j$ of the jet material
is of 150~km~s$^{-1}$
at $x=0$ (correctly coinciding with the injection velocity,
see Table~1). For model a1 (top panel of Figure~\ref{fig6}), the
jet velocity drops close to the position of
the head to $\sim 120$~km~s$^{-1}$.
The drop in $\overline{v}_j$ occurs closer to the jet source
in model c1 (bottom panel of Figure~\ref{fig6}). 
In both models (a1 and c1), the mean forward velocity
$\overline{v}_{AM}$ of the 
environmental material grows along
the length of the outflow, with a velocity of $\approx 120$~km~s$^{-1}$
at the jet head. At all times, we also see a peak in $\overline{v}_{AM}$
at $x<10^{16}$~cm (the position of this peak being at larger distances
from the source for longer integration times). This peak is associated
with a region (close to the jet source) in which the post-bow shock
shell touches the jet beam since early evolutionary times
(see Figure~\ref{fig2}).

\begin{figure}[!h]
\epsscale{1.1}
\plotone{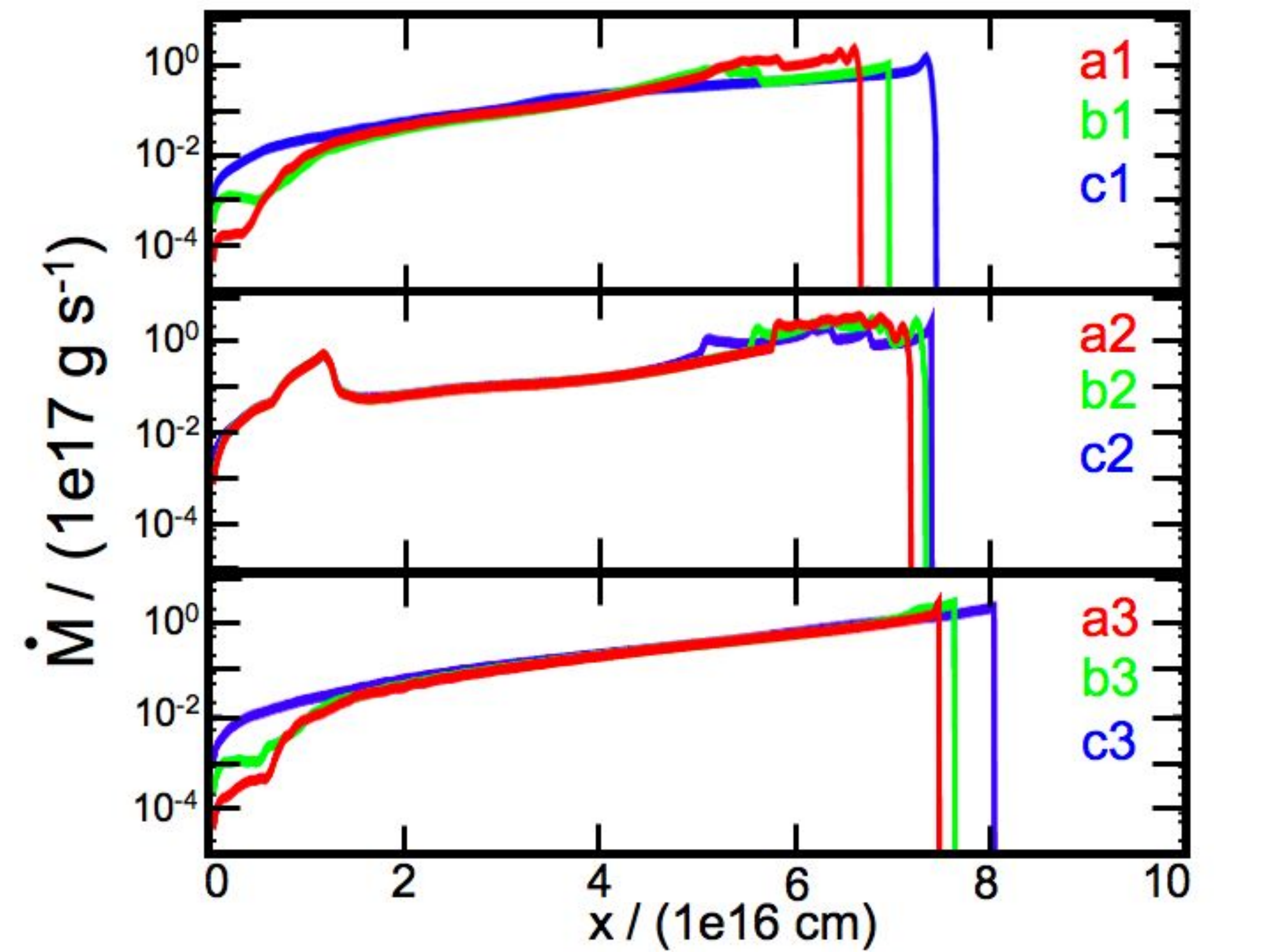}
\caption{Environmental mass rate ${\dot M}_{AM}$ (see equation
\ref{mam}) computed for all models (see Table 1)
for an integration time of 200~yr for models a1, b1,
c1, a3, b3 and c3 and of 100~yr for models a2, b2 and c2.}
\label{fig7}
\end{figure}

In Figure~\ref{fig7}, we show the environmental mass rate
as a function of distance from the source obtained
from all of the models of Table~1 (only the results
for a single integration time are shown for each model).
All of the models give ${\dot M}_{AM}\approx
10^{17}$~g~s$^{-1}$ close to the head of the jet.
Lower mass loss rates are obtained closer to the
outflow source.

The models with $v_j=300$~km~s$^{-1}$ (a2, b2 and c2, central panel) show basically identical ${\dot M}_{AM}(x)$ dependencies for $x<5\times 10^{16}$~cm, and a clear spike at $x_{sp} \approx 10^{16}$~cm, but as we shall see corresponds to environmental material which is not entrained primarily by the sidewind, but rather from the leading bow-shock, thus is not of interest for this study. The $v_j=150$~km~s$^{-1}$ models (top and bottom panels) show a region close to the outflow source (with $x<2\times 10^{16}$~cm) in which larger ${\dot M}_{AM}$ are obtained for larger sidewind velocities ($v_a=2$~km~s$^{-1}$ for models a1 and a3, 5~km~s$^{-1}$ for b1 and b3 and 10~km~s$^{-1}$ for c1 and c3, see Table 1). As we will see in the following section, in this region we are seeing environmental material that is directly entrained into the jet beam (due to the presence of the sidewind).

\begin{figure}[!h]
\epsscale{1.1}
\plotone{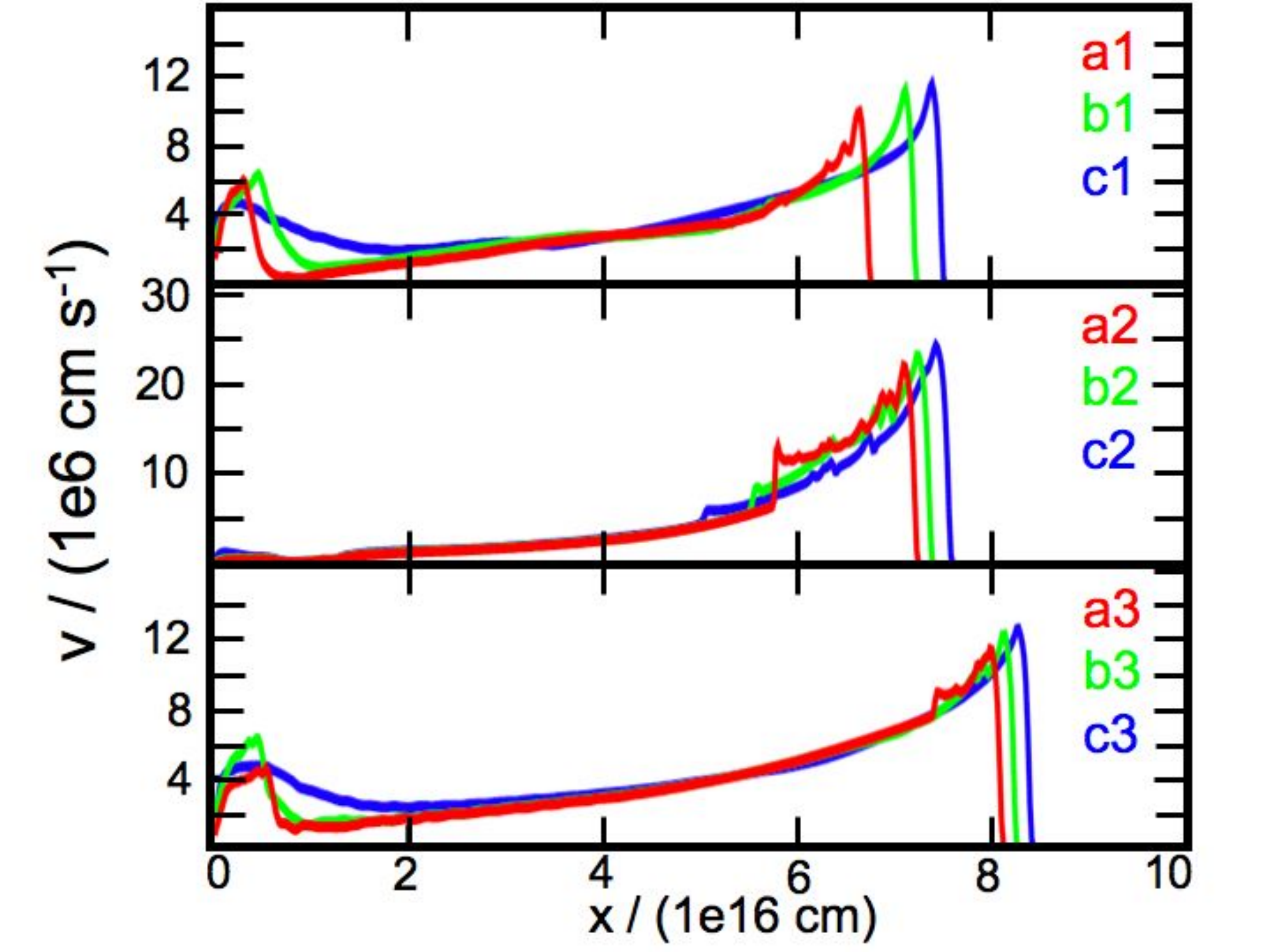}
\caption{Environmental average velocity ($\overline{v}_{AM}$) (see equation
\ref{vam}) computed for all models (see Table 1)
for an integration time of 200~yr for models a1, b1,
c1, a3, b3 and c3 and of 100~yr for models a2, b2 and c2.}
\label{fig8}
\end{figure}

Figure~\ref{fig8} shows the environmental mean velocity ($\overline{v}_{AM}$, see equation \ref{vam}), as a function of position for all of the computed models (see Table 1). In the $v_j=150$~km~s$^{-1}$ models (upper and lower panels of Figure~\ref{fig8}), regardless of the sidewind velocity the correspondent $\overline{v}_{AM}$ grows from low ($\sim 1$~km~s$^{-1}$) close to the source, up to values comparable to the jet velocity ($\sim 250$~km~s$^{-1}$) at the head of the jet.

Very similar results are obtained for the models with $v_j=300$~km~s$^{-1}$ (central panel), except for the presence of a spike at $x_{sp} \approx 10^{16}$~cm. Since the corresponding $\overline{v}_{AM}$ at $x_{sp}$ for these models ($\sim 0.03 v_j$) was lower than that from the models with $v_j=150$~km~s$^{-1}$ (where $\overline{v}_{AM} \sim 0.04 - 0.20 v_j$), we had an insight that the material at the spike of Figure~\ref{fig7} (for models with $v_j=300$~km~s$^{-1}$) corresponded to slowly moving material which was mostly entrained by the leading bow-shock and not due to the presence of the sidewind. 

Also in Figure~\ref{fig8}, there is evidence of ``high velocity materialÓ. Our models with $v_j=150$~km~s$^{-1}$ showed the presence of a fast region close to the source ($x<2\times 10^{16}$~cm). In this region $\overline{v}_{AM}$ has values of $\sim 40\to 70$~km~s$^{-1}$ (of order $\sim 50$\%\ of the $v_j$). These large $\overline{v}_{AM}$ values close to the source are associated with the direct entrainment of the dense post-bow shock shell into the jet beam due to the sidewind.

\section{Fast entrained material}
\label{sec:fast}

In order to study the environmental material which
is directly entrained into the jet beam, we compute
the mass rate and average velocity of the environmental
material that moves along the jet axis at velocities
$>v_j/2$ (in other words, with velocities larger than
75~km~s$^{-1}$ for models a1, b1, c1, a3, b3, c3 and
larger than 150~km~s$^{-1}$ for models a2, b2 and c2).
We call these the ``high velocity'' mass rate ${\dot M}_{AM,h}$
and average velocity $\overline{v}_{AM,h}$.

\begin{figure}[!h]
\epsscale{1.1}
\plotone{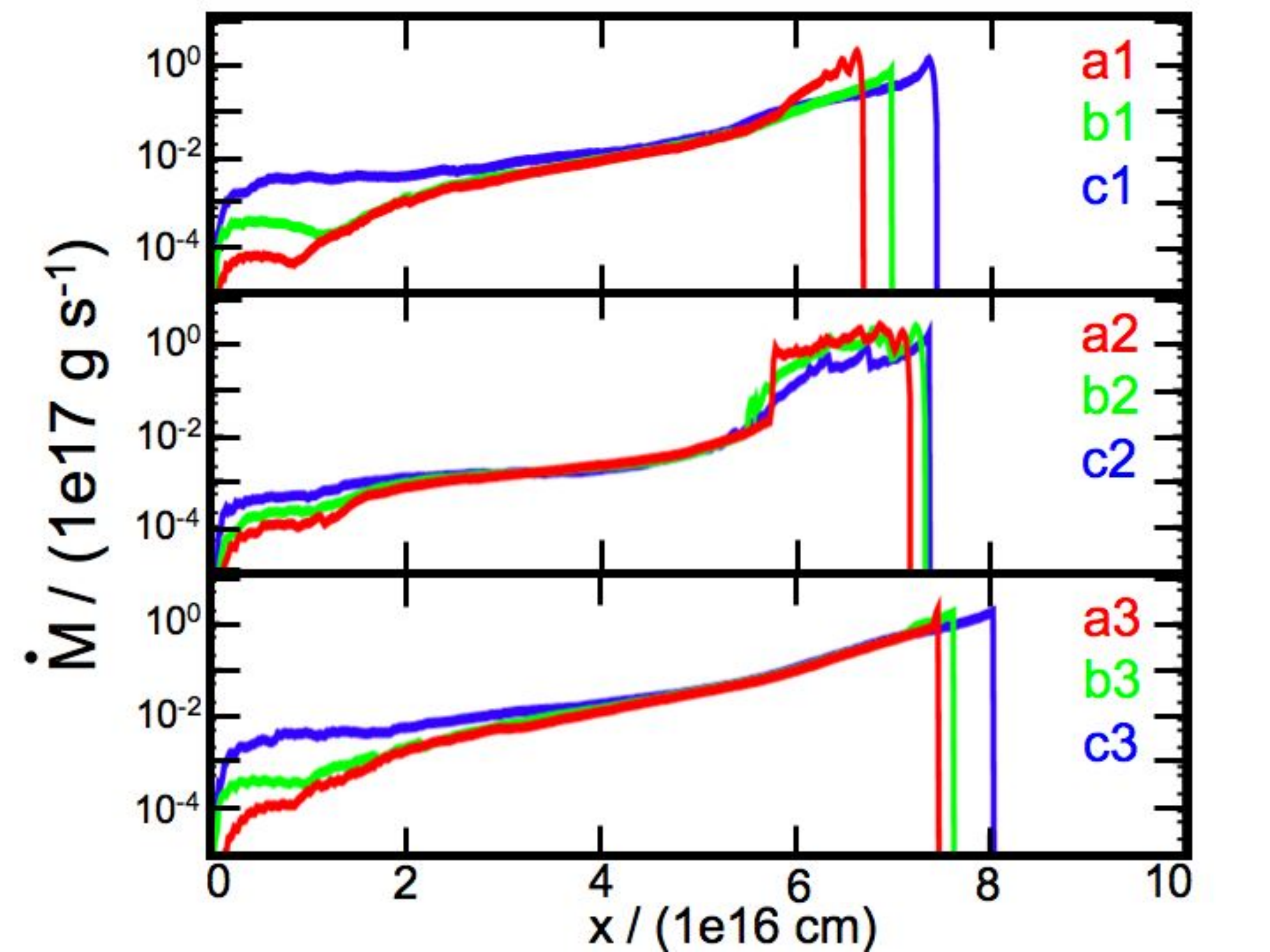}
\caption{Mass rate of the environmental
material which moves with ``high" velocities along the jet axis (${\dot M}_{AM,h}$). For model a1, b1, c1, a3, b3, c3 we show the material that is moving with $\overline{v}_{AM}>$75~km~s$^{-1}$, and with $\overline{v}_{AM}>$150~km~s$^{-1}$ for models a2, b2 and c2.}
\label{fig9}
\end{figure}

The values of ${\dot M}_{AM,h}(x)$ obtained for all models are shown in Figure~\ref{fig9}. We see that in the region close to the outflow source ($x<2\times 10^{16}$~cm) the low velocity jet models (with $v_j=150$~km~s$^{-1}$, models a1, b1, c1, a3, b3, c3) have environmental mass rates which monotonically grow with increasing values of the sidewind velocity ($v_a$). For
$v_a=10$~km~s$^{-1}$ (models c1 and c3), the mass rate in this region has values of ${\dot M}_{AM,h} \approx 5\times 10^{14}$~g~s$^{-1}$, of the order of $\sim 0.5$\%\ of the mass loss rate of the jet. Lower values of ${\dot M}_{AM,h}$ are obtained for the $v_a=5$ (b1 and b3) and $v_a=2$~km~s$^{-1}$ models (a1 and a3).
Finally, we see that the values of ${\dot M}_{AM,h}$ in the region close to the source are much lower (${\dot M}_{AM,h} \approx 5\times 10^{13}$~g~s$^{-1}$) for the $v_j=300$~km~s$^{-1}$ models (a2, b2, c2, central panel of Figure~\ref{fig9}). This confirms the fact that the material from the spike in Figure~\ref{fig7} (for models with $v_j=150$~km~s$^{-1}$), corresponds to environmental material which was not entrained in its majority by the sidewind (and so, is not of interest for this study).

All of the models show a strongly increasing ${\dot M}_{AM,h}(x)$ for
$x>2\times 10^{16}$~cm. This growth is associated with the fact
that the motion of the post-bow shock shell becomes progressively
more forward directed as we approach the head of the jet.

\begin{figure}[!h]
\epsscale{1.1}
\plotone{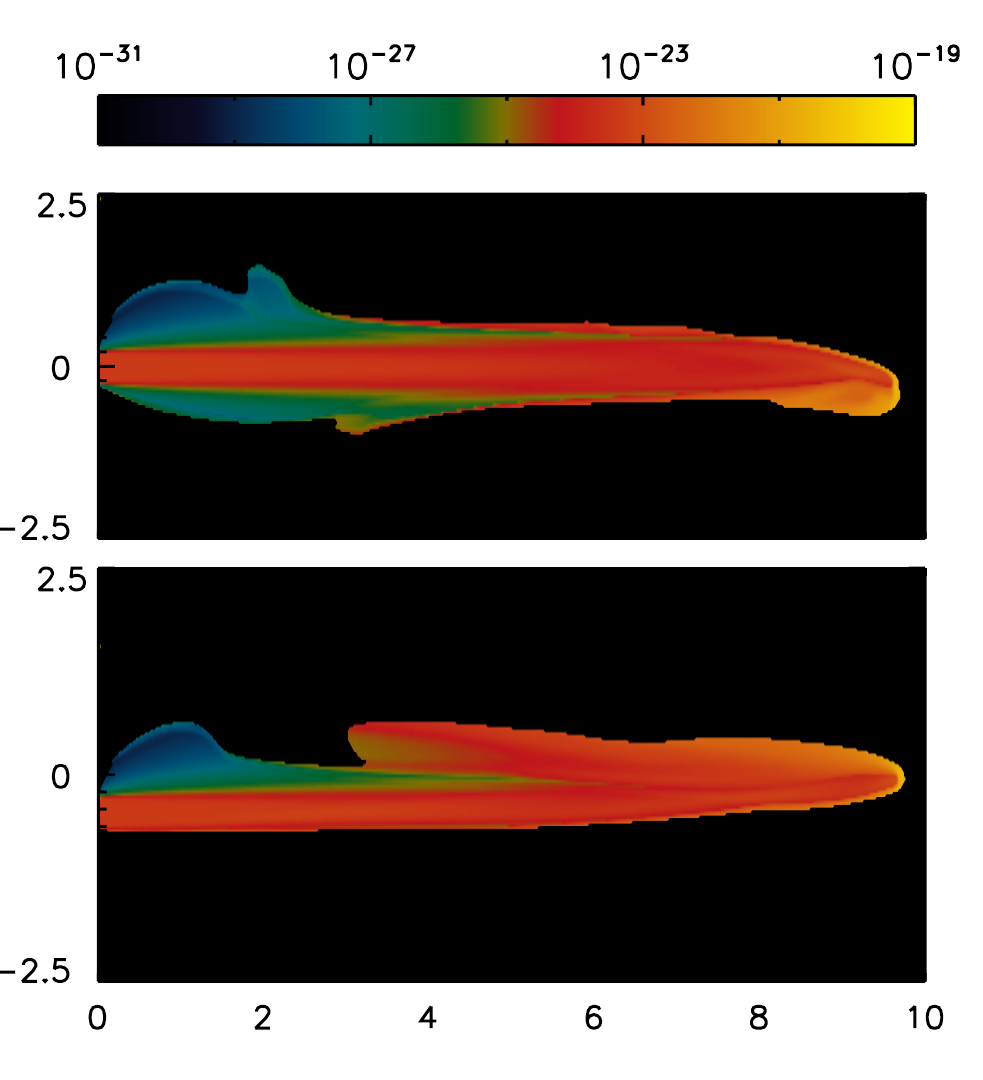}
\caption{Environmental mass fraction times the density stratification ($\rho \times f_m$, colour scale, given in g cm$^{-3}$ by the top bar), of the fast ambient material obtained from models a1 (top panel) and c1 (bottom panel), for an integration time of 300~yr. The axes are labeled in units of $10^{16}$~cm.}
\label{fig10}
\end{figure}

Therefore, there are two components of the fast (with
axial velocities $>v_j/2$) ambient material~:
\begin{itemize}
\item material originating in the region in which
the post-bow shock shell is in contact with the jet
beam,
\item material in the post-bow shock shell in the
region close to the jet head.
\end{itemize}
The spatial distribution of these two components is
shown in Figure~\ref{fig10}, with the mid-plane spatial distribution
of the density $\rho f_m$ of the fast ($v_x>v_j/2$)
ambient material obtained
from models a1 and c1 (for an integration time of 300~yr).
These stratifications show that the fast ambient material is
confined to a region within or in near contact to the jet beam.

\section{A resolution study}
\label{sec:res}

\begin{figure}[!h]
\epsscale{1.25}
\plotone{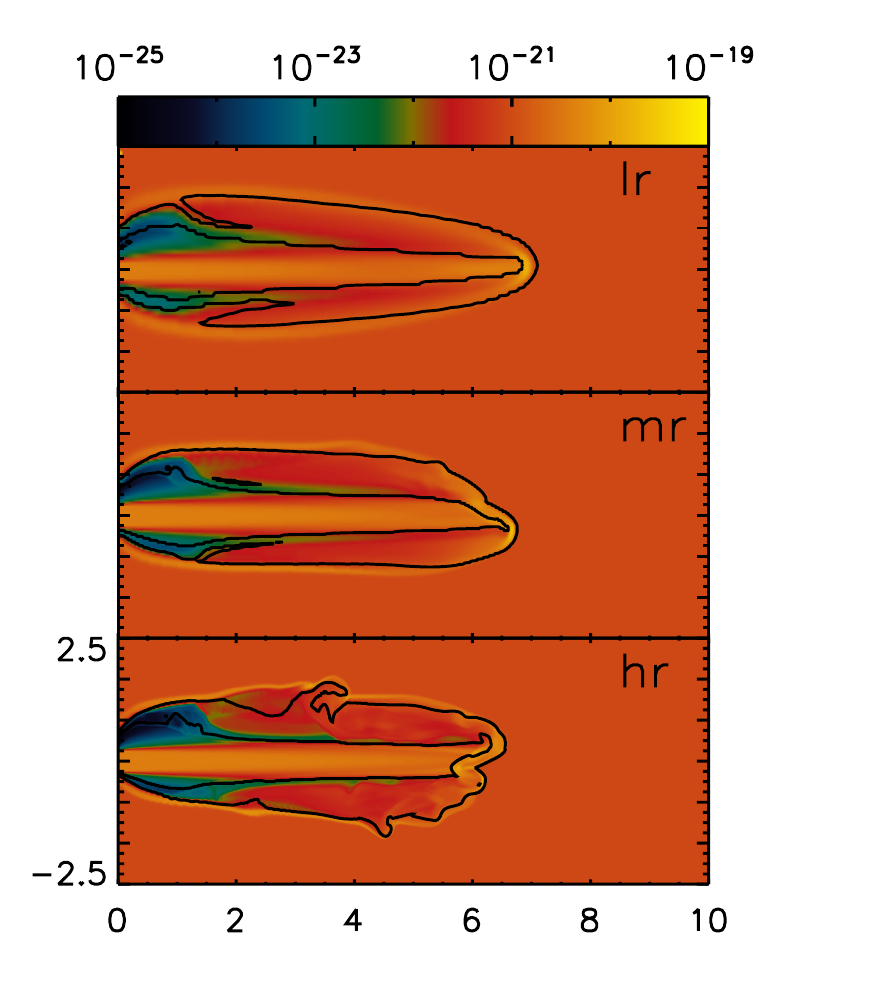}
\caption{Density stratifications (color scale, given in g cm$^{-3}$ by the top bar) and mixing fractions (contours) obtained from model a1 at three different resolutions for an integration time of 200~yr (see Table 1). The two contours correspond to the 99~\%\ mixing fraction ($f_m=0.99$, outer contour); and the 1~\%\ mixing fraction case ($f_m=0.01$, inner contour). The axes are labeled in units of $10^{16}$~cm. The three panels correspond to ``low" (case in which $\Delta x = 3.91\times 10^{13}$~cm along the three axes, top panel), ``medium" ($\Delta x = 1.95\times 10^{13}$~cm), and ``high" ($\Delta x = 0.98\times 10^{13}$~cm, bottom panel) resolutions.}
\label{fig11}
\end{figure}

In order to illustrate the effect of the numerical resolution,
we have computed two of the models (a1 and c1, see Table 1)
at three resolutions: $\Delta x = 3.91$, 1.95 and
$0.98\times 10^{13}$~cm (along the three axes). In Figure~\ref{fig11},
we show the density stratifications obtained from model a1
at these three resolutions, for a $t=200$~yr integration time.

It is clear that more complex structures are obtained
for increasing resolutions. However, the result
that the post-bow shock shell is swept into contact
with the jet beam (as a result of the side-streaming
environment) is present at all resolutions (see Figure~\ref{fig11}).
Therefore, side-entrainment into the jet beam occurs regardless
of the resolution of the simulations.

\begin{figure}[!h]
\epsscale{1.2}
\plotone{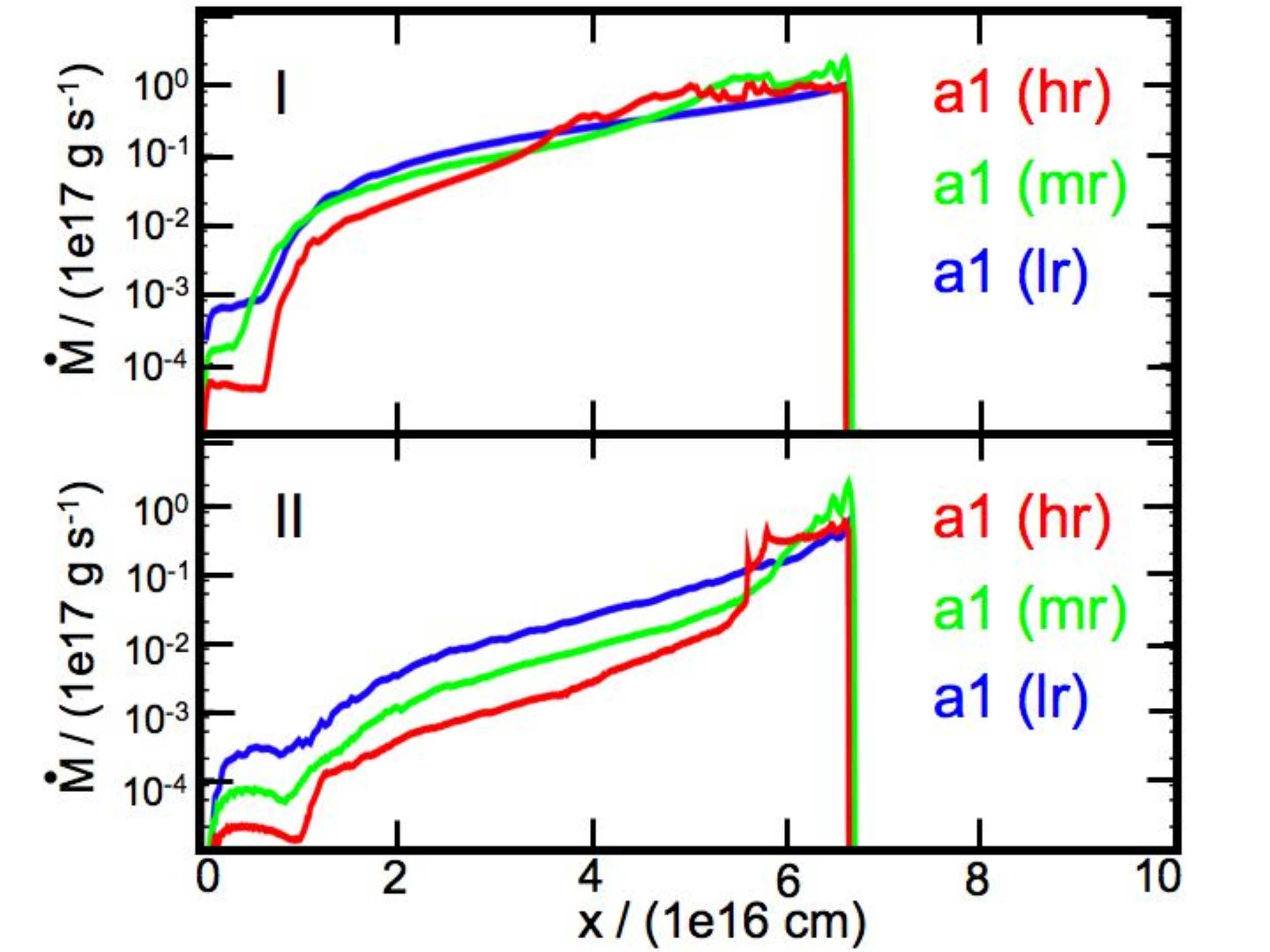}
\caption{Total environmental mass rate (${\dot M}_{AM}$, top panel) and the mass rate of the fast entrained material (${\dot M}_{AM,h}$, bottom panel), as a function of distance from the source, computed for the three different resolutions from model a1 (at an integration time $t=200$~yr).}
\label{fig12}
\end{figure}

In Figure~\ref{fig12}, we show the total environmental mass
rate (${\dot M}_{AM}$, top frame) and the mass rate of the fast
entrained material (${\dot M}_{AM,h}$, bottom frame) as
a function of distance from the source, computed from
the $t=200$~yr stratification of model a1.
The values of ${\dot M}_{AM}$ and ${\dot M}_{AM,h}$
are similar at all resolutions in the region close
to the jet head.

On the other hand, in the region close
to the source (where the dense, post-bow shock shell is
being entrained into the jet beam), the environmental
mass rate shows a stronger dependence on the resolution.
In particular, we have a factor of $\sim 10$ decrease in 
${\dot M}_{AM,h}$ for a factor of 4 increase in the resolution
of the simulation. This result is to be expected (given the
lower numerical diffusion of the higher resolution simulations),
and its implications are discussed in the following section.

It is clear that in this resolution study a relatively small range (of a single octave) of resolutions is explored. This is a result of the fact that simulations with less than $\sim 10$ grid points across the jet beam are basically meaningless, and that our highest resolution simulation (model hr) resolves the jet beam diameter with $\sim 100$ points (at the highest resolution of the adaptive grid). This resolution is ``competitive'' in terms of recent 3D astrophysical jet simulations, for example \citet{rossi08} computed 3D simulations of entrainment in relativistic jets resolving the jet diameter with $\sim 20$ points).
In the near future we will be able to carry out 3D simulations with higher resolutions, hopefully reaching $\sim 300-500$ points across the jet diameter. This will extend the range in resolutions to $\sim 1.5$ octaves (which is still not very impressive).

The situation is therefore somewhat hopeless. If we take the number of grid points across the jet diameter as an estimate of the Reynolds number of the simulation, we see that at the present time we can only expect to reach Re$\sim 500$. Such Reynolds numbers are $\sim 2$ orders of magnitude below the Reynolds number (of $\sim 50000$) necessary to reach the ``high Re" regime, in which the entrainment rate becomes independent of Re \citep{be72}. Therefore, it is unclear if this ``high Re" regime (relevant for astrophysical jets) will ever be reached by 3D gasdynamic simulations. Regardless of this fact, exploring the development of the Kelvin-Helmholtz instabilities (that initiate the turbulent entrainment) at increasing spatial resolutions is probably still a worthwhile pursuit. Different efforts in this direction have been done by \citet{micono00a, micono00b, xu00, rh02}.

An eventually more fruitful approach might be to incorporate a ``turbulence parametrization scheme", which has to be calibrated with laboratory experiments in order to provide the correct mass entrainment rate. The application to astrophysical jets of a simple, ``$\alpha$-model" turbulence parametrization was discussed by \citet{can91}. A more evolved, ``k-$\epsilon$" scheme was explored by \citet{fal94}.

The problem of this approach of course is that the turbulence parametrization schemes are normally calibrated with non-radiative, laboratory jet flow experiments. Therefore, it is not clear whether or not the resulting parametrization is appropriate for modeling the entrainment in radiative jets. A possible solution to this problem would be to attempt to calibrate turbulence parametrization schemes with newly available experiments of radiative plasma jets \citep[see, e.~g.][]{ample08}. The relevant experimental data, however, are still not available.

\section{Conclusions}
\label{sec:conc}

It has been suggested that the molecular emission observed
along well collimated, jet-like outflows from young stars
might be the result of entrainment of molecular environmental
material into the jet beam \citep{can91}. However,
this side-entrainment has never
been seen in numerical simulations of HH jets, because
the environmental material is pushed into a dense, post-bow
shock shell which does not touch the jet beam.

In this paper, we present 3D numerical simulations of a jet
in a sidewind, with sidewind velocities in the $v_a=2\to 10$~km~s$^{-1}$
range (the lower part of this range being consistent with the
peculiar motions of T~Tauri stars). We find that for many parameter
combinations the sidewind pushes the post-bow shock shell into direct
contact with the jet beam (see Figure~\ref{fig2} and \ref{fig3}).
In this region of contact, side-entrainment
of environmental material into the jet beam does take place.

In our simulations, the side-entrainment results in mass rates
${\dot M}_{AM,h}\sim 5\times 10^{14}$~g~s$^{-1}$, corresponding
to $\sim 0.5$~\%\  of the mass loss rate ${\dot M}_j$ of the jet
(see \S\ref{sec:ent}). If the molecular, environmental gas
is not dissociated during the process of side-entrainment,
this would result in a molecular fraction of $\sim 0.5$~\%\ 
within the jet beam, which would result in molecular column
densities high enough to produce observable molecular
emission \citep{rag05}.

Our present simulations do not include the chemistry of
the entrained material, so that we are not able to see
whether or not the molecules in the side-entrained
material actually survive the entrainment process.
However, the fact that the side-entrained material has been
shocked by the slow-moving far bow shock wings, and that
the region of contact between the shocked environment
and the jet remains cool (at temperatures of $\sim 10^3$~K)
indicates that molecules
indeed might be entrained into the jet beam without being
dissociated.

Furthermore, our simulations do not describe correctly the
entrainment of the post-bow shock shell into the jet beam.
An indication of this is the fact that we obtain entrained
mass rates that strongly depend on the spatial resolution
of our simulations (see \S\ref{sec:res}). In order to overcome this
problem, it will be necessary to go to much higher resolutions,
in order to resolve the Kelvin-Helmholtz instabilities that
produce the side-entrainment \citep[see e.~g][]{micono00a, micono00b},
or to use a ``turbulence parametrization recipe'' \citep{can91,fal94}.

Even though our simulations do not fully describe the
side-entrainment process, they conclusively show that a
side-streaming environment (reflecting the
motion of the jet source) will push the post-bow shock shell
into direct contact with the jet beam. This then provides the
conditions in which molecular material will be entrained into the
fast jet beam, giving ``parmetrized turbulence'' jet models -initially suggested for astrophysical jets by \citet{bic84}-, a new life as a possible explanation for the
molecular jets observed in star forming regions.

\acknowledgements
This work was supported by the CONACyT grants 61547 and 101356.

\end{document}